\documentclass[aps,prl,nofootinbib,12pt]{revtex4-1}

\usepackage{amssymb,graphics,graphicx}
\usepackage{footnote}
\usepackage{fullpage}

\usepackage{amsmath, amsfonts}

\begin{document}

\title
{\Large \bf Remarks on the Cosmological Constant}

\author{Paul H. Frampton\footnote{frampton@physics.unc.edu}}

\affiliation{Department of Physics and Astronomy, University of North Carolina,\\
Chapel Hill, NC 27599-3255, USA\footnote{Permanent address}\\
and\\
Centro Universitario Devoto, Buenos Aires, Argentina}

\date{\today}
\begin{abstract}
\begin{center} \textbf{Abstract} \end{center}
The acceleration of the surface of last scatter (SLS) must somehow reflect the energy content within it.  A test particle at the SLS is assumed to experience 
a linear combination of two forces:  one Newtonian, the other pseudo-Newtonian describable by a cosmological constant ${\Lambda}$ in general relativity.  In 
the ${\Lambda}$ description, which is surely too unimaginative, the size of ${\Lambda}$ reflects only the comparable magnitudes of the Newtonian and pseudo-Newtonian forces; any claim of fine tuning due to quantum mechanics is probably illusory.
\end{abstract}

\maketitle

\newpage
The two most groundbreaking discoveries made in observational cosmology are closely related: the expanding universe was convincingly established by Hubble in 1929
and the accelerating nature of this expansion by Perlmutter, Riess and Schmidt \cite{perlmutter,kirshner} in 1998 deservedly honored by the 2011 Nobel prize.

The Hubble expansion could have been, but was not, predicted by Einstein from general relativity in 1917, had he taken the simplest stable solution more seriously and confidently. Here we shall discuss the acceleration, and associated dark energy, discovered sixty-nine years later.  

On an alternative language, we shall argue the cosmological constant is necessarily non-zero and positive, and has the value $\Lambda \sim 10^{-124}$ in Planck units, although this apparent fine tuning is, as we shall argue, surely illusory, and the expansion of the universe must therefore be accelerating.

To flesh out these general remarks, we will cite as data principally the precision results of WMAP7 \cite{komatsu} and the more global averages cited therein. It seems appropriate to begin with the present value of the Hubble constant $\mathcal{H}(t_0)$ which is given by\footnote{The central value is quoted, without errors, since the present article addresses only a conceptual issue.}
\begin{eqnarray}\label{eq:H(t_0)}
\mathcal{H}(t_0) &=& 71 \, \rm{km}/\rm{s}/\rm{Mpc}\\
&=& 2.30 \times 10^{-18} \, \rm{s}^{-1}
\end{eqnarray}
accurate to $5\%$. Using for Newton's constant $G = 6.671 \times 10^{-11} \, \rm{m}^3 \rm{kg}^{-1} \rm{s}^{-2}$ this leads to a value for the critical density of
\begin{eqnarray}
\rho_c(t_0) &=& \frac{3 \mathcal{H}(t_0)^2}{8 \pi G}\\
&=& 9.46 \times 10^{-27} \, \rm{kg}\, \rm{m}^{-3}
\end{eqnarray}
again giving only a central value, accurate to $10\%$.

The co-moving radius $\mathcal{R}(t_0)$ of the electromagnetically visible universe, out to the surface of last scatter (SLS) for photons, is known with extraordinary precision. General relativity is used in defining $\mathcal{R}(t_0)$ in terms of the scale factor $a(t)$ in a Friedmann-Lema\^{\i}tre-Robertson-Walker (FLRW) metric
\begin{equation}
\mathcal{R}(t_0) = C \int_0^{t_0} \frac{dt}{a(t)}
\end{equation}
and is given in \cite{komatsu} by
\begin{eqnarray}
\mathcal{R}(t_0) &=& 46.0 \rm{Gly}\\
&=& 4.36 \times 10^{26} \rm{m}\label{eq:R(t_0)}
\end{eqnarray}
with accuracy better than $1\%$.

Thus the visible universe so defined as a sphere characterized by Eqs. (\ref{eq:H(t_0)}-\ref{eq:R(t_0)}), although otherwise unimaginable large can be studied assiduously with precision quite comparable to measuring a human-size object like a basket ball by using a bathroom scale and a tape measure.

The total mass-energy $\mathcal{M}(t_0)$ inside the SLS is calculable assuming vanishing curvature to be, accurate to $2\%$,
\begin{eqnarray}
\mathcal{M}(t_0) &=& \tfrac{4 \pi}{3} \rho_c(t_0) \mathcal{R}(t_0)^3\\
&=& 3.28 \times 10^{54} \, \rm{kg}\,.
\end{eqnarray}

This total mass-energy $\mathcal{M}(t_0)$ is traditionally divided into three pieces: baryonic matter, non-baryonic dark matter and dark energy which is the source of the accelerated expansion rate.

From $\mathcal{M}(t_0)$ and $\mathcal{R}(t_0)$ we may compute the Newtonian gravitational acceleration $\kappa$ at the SLS from
\begin{eqnarray}
\kappa(t_0) &=& \frac{G \mathcal{M}(t_0)}{\mathcal{R}(t_0)^2}\\
&=& 1.15 \times 10^{-9} \, \rm{m}\, \rm{s}^{-2}\label{eq:kappa}
\end{eqnarray}
so that for a test object at the SLS the gravitational force of attraction towards the center is $F = m \kappa$.

In general relativity, by contrast, this force $F$ does not exist. Gravity is instead a result of spacetime curvature.

Eq. (\ref{eq:kappa}) treats all the energy in the visible universe equally, and the result disagrees with a sign and slightly with the magnitude found by observation of 
supernovae \cite{perlmutter,kirshner}, which instead suggest the $16\%$ larger magnitude:

\begin{eqnarray}
\kappa_{obs}(t_0)=-1.34 \times 10^{-9} \, \rm{m}\, \rm{s}^{-2}. \label{eq:kappa_obs} 
\end{eqnarray}
which is close in size to $\kappa(t_0)$ by virtue of the approximate conicidence between matter and dark energy.

Taking only the matter component, $\Omega_m = 0.28$, the partial acceleration analogous to Eq. (\ref{eq:kappa}) is:

\begin{eqnarray}
\kappa_m(t_0) = \Omega_m \kappa(t_0)= 3.3 \times 10^{-10} \, \rm{m}\, \rm{s}^{-2}.\label{eq:kappa_m}
\end{eqnarray}

For the $\Omega_{\Lambda} = 0.72$ component, adopting an unknown factor $\eta$ to correct the force with the test particle gives:

\begin{eqnarray}
\kappa_{\Lambda}(t_0)=\eta \Omega_{\Lambda} \kappa(t_0) \label{eq:kappa_Lambda}
\end{eqnarray}
as corresponds to a gravity law $S=\eta S_{Newton}$.  

The overall inward acceleration is then:

\begin{eqnarray}
\kappa_{TOT}(t_0)=\kappa_{m}(t_0) + \kappa_{\Lambda}(t_0) \\
=(0.33 + 0.96 \eta) \times 10^{-9} \, \rm{m}\, \rm{s}^{-2}. \label{eq:kappa_TOT}
\end{eqnarray}
If Eq. (\ref{eq:kappa}) and Eq. (\ref{eq:kappa_obs}) had equal magnitudes, then $\eta \simeq -1.8$.  The observed result, Eq. (\ref{eq:kappa_obs}), requires instead $\eta \simeq -2.0$, very closely \cite{wald}.

Acceleration of the FLRW scale factor is:
\begin{eqnarray}
\ddot{a}(t_0) = -\kappa_{TOT}(t_0) \mathcal{R}(t_0)^{-1} = 3.1 \times 10^{-36} \, \rm{s}^{-2}.
\end{eqnarray}

Written as an energy density this gives
\begin{equation}
\rho_\Lambda(t_0) \sim + 10^{-26} \, \rm{kg}/\rm{m}^3.
\end{equation}

In Eq. (\ref{eq:kappa_Lambda}), the curious factor $\eta \simeq -2.0$ can be reproduced in general relativity with non-zero $\Lambda$, but general relativity is untested at 
scales $\mathcal{R}(t_0) \sim 10^{26} \rm{m}$, only for much smaller scales up to $\lesssim 10^{12} m$, so since the cosmological constant $\Lambda$, or whatever may replace 
it, dominates the fate and possible cyclicity of the universe, it cries out for a more satisfactory theory of $\eta \simeq -2.0$ than the $\Lambda$CDM parameterization.

It is sometimes claimed that another difficulty is that one ``expects''
\begin{equation}
\rho_\Lambda(t_0) \sim m_{\rm{Pl}} / l_{\rm{Pl}}^3 \simeq 10^{98} \, \rm{kg} / m^3
\end{equation}
if, as might be hinted by quantum mechanics, the naturally occurring scale is the Planck scale. We do not encounter such an argument which invokes Planck's constant that plays no role in the appearance of non-zero $\Lambda$.

Thus there is absolutely no reason to expect $\rho_\Lambda(t_0)$ to be $10^{124}$ times bigger than it is known to be. The difference arises only because $\mathcal{R}(t_0)$ is $10^{62}$ times the Planck length $l_{\rm{Pl}} \sim 10^{-35} \, m$ while $\mathcal{M}(t_0)$ is, coincidentally, $10^{62}$ times the Planck mass $m_{\rm{Pl}} \sim 10^{-9} \, \rm{kg}$; so $\mathcal{M}(t_0)/\mathcal{R}(t_0)^3$ is $10^{-124}$ times the Planck value. A Planck-type universe with the observed total mass-energy $\mathcal{M}(t_0)$ would have the radius of an atomic nucleus: one with the observed radius $\mathcal{R}(t_0)$ would have $10^{124}$ as many stars as exist.

We may expect that the pseudo-Newtonian force with $\eta \simeq -2.0$ will become better understood by study of the ideas based on thermodynamics and entropy 
discussed in \cite{easson,verlinde}.  Another fruitful direction can be provided by the observation that the values of 
$\mathcal{M}(t_0)$ and $\mathcal{R}(t_0)$ discussed at {\it ut supra} 
imply that the SLS closely approximates the surface of a gigantic black hole, as discussed in \cite{gell}.

~

\noindent{\bf Acknowledgements:}
I am grateful to K. Ludwick for a useful discussion and to the Centro Universitario Devoto for providing a generous amount of unencumbered time. This work was supported in part by U.S. Department of Energy Grant No. DE-FG02-05ER41418.  


\end{document}